\documentclass[a4paper,fleqn,usenatbib]{mnras}

\usepackage{mathptmx}

\usepackage[T1]{fontenc}
\usepackage{ae,aecompl}

\usepackage{graphicx}	
\usepackage{amsmath}	
\usepackage{amssymb}	
\usepackage{mathrsfs}

\title[GW150914-like BHs as Galactic High-Energy Sources]
{GW150914-like Black Holes as Galactic High-Energy Sources}

\author[K. Ioka et al.]{
Kunihito Ioka,$^{1}$\thanks{E-mail: kunihito.ioka@yukawa.kyoto-u.ac.jp (KI)}
Tatsuya Matsumoto,$^{2}$
Yuto Teraki,$^{1}$
Kazumi Kashiyama$^{3}$
and \newauthor
Kohta Murase$^{1,4,5,6}$
\\
$^{1}$Center for Gravitational Physics, Yukawa Institute for Theoretical Physics, Kyoto University, Kyoto 606-8502, Japan\\
$^{2}$Department of Physics, Graduate School of Science, Kyoto University, Kyoto 606-8502, Japan\\
$^{3}$Department of Physics, University of Tokyo, Bunkyo, Tokyo 113-0033, Japan\\
$^{4}$Department of Physics, The Pennsylvania State University, University Park, PA 16802, USA\\
$^{5}$Department of Astronomy \& Astrophysics, The Pennsylvania State University, University Park, PA 16802, USA\\
$^{6}$Center for Particle and Gravitational Astrophysics, The Pennsylvania State University, University Park, PA 16802, USA\\
}

\date{Accepted XXX. Received YYY; in original form ZZZ}

\pubyear{2016}

\begin{document}
\label{firstpage}
\pagerange{\pageref{firstpage}--\pageref{lastpage}}
\maketitle

\begin{abstract}
The first direct detections of gravitational waves (GWs) 
from black hole (BH) mergers, GW150914, GW151226 and LVT151012,
give a robust lower limit $\sim 70000$ on the number of 
merged, highly-spinning BHs in our Galaxy.
The total spin energy is comparable to all the kinetic energy of 
supernovae that ever happened in our Galaxy.
The BHs release the spin energy to relativistic jets
by accreting matter and magnetic fields from the interstellar medium (ISM).
By considering the distributions of the ISM density, BH mass and velocity, 
we calculate the luminosity function of the BH jets,
and find that they can potentially accelerate
TeV--PeV cosmic-ray particles in our Galaxy with total power
$\sim 10^{37\pm 3}$ erg s$^{-1}$
as PeVatrons, positron factories and/or
unidentified TeV gamma-ray sources.
Additional $\sim 300$ BH jet nebulae could be
detectable by CTA (Cherenkov Telescope Array).
We also argue that the accretion from the ISM 
can evaporate and blow away cold material around the BH,
which has profound implications for some scenarios 
to predict electromagnetic counterparts to BH mergers.
\end{abstract}




\section{Introduction}

A century after Einstein predicted the existence of gravitational waves (GWs),
the Laser Interferometer Gravitational-Wave Observatory (LIGO)
observed the first direct GW signal GW150914
from a merger of two black holes (BHs)
with masses of $36_{-4}^{+5} M_{\odot}$ and $29_{-4}^{+4} M_{\odot}$
and radiated energy $3_{-0.4}^{+0.5} M_{\odot} c^2$.
\citep{LIGO_1st}.
This is also the first discovery of a binary BH.
During Advanced LIGO's first observing period (O1),
September 12, 2015 to January 19, 2016 \citep{LIGO_O1},\footnote{
O1 was officially September 18, 2015 to January 12, 2016 
before the detection of GW150914.}
the second event GW151226 with masses
$14.2_{-3.7}^{+8.3} M_{\odot}$ and $7.5_{-2.3}^{+2.3} M_{\odot}$
and radiated energy $1.0_{-0.2}^{+0.1} M_{\odot} c^2$
\citep{LIGO_GW151226}
and a candidate LVT151012
with $23_{-6}^{+18} M_{\odot}$, $13_{-5}^{+4} M_{\odot}$
and $1.5_{-0.4}^{+0.3} M_{\odot} c^2$
have been also detected,
and the existence of a population of merging BHs has been established.
These $\sim 2.5$ events give a relatively certain estimate
on the merger rate in the range
$\mathscr{R}_{\rm GW} \sim 9$--$240$ Gpc$^{-3}$ yr$^{-1}$ \citep{LIGO_O1}.
A new era of GW astrophysics has been finally opened
and will be driven by a network of LIGO, Virgo, KAGRA, and IndiGO,
and by eLISA and DECIGO satellites in the future
\citep{Sesana16,Kyutoku_Seto16,Nakamura+16b}.

The binary BH mergers are the most luminous events in the universe,
even brighter than gamma-ray bursts (GRBs).
The peak luminosities are $3.6_{-0.4}^{+0.5} \times 10^{56}$ erg s$^{-1}$,
$3.3_{-1.6}^{+0.8} \times 10^{56}$ erg s$^{-1}$
and $3.1_{-1.8}^{+0.8} \times 10^{56}$ erg s$^{-1}$
for GW150914, GW151226 and LVT 151012, respectively \citep{LIGO_O1},
which reach $\sim 0.1\%$ of the Planck luminosity
$c^5/G=3.6 \times 10^{59}$ erg s$^{-1}=2.0 \times 10^{5} M_{\odot} c^2$ s$^{-1}$.
Merged BHs also retain huge energy in the spin.
The spin energy is about
\begin{eqnarray}
E_{\rm spin} &=& \left(1-\sqrt{\frac{1+\sqrt{1-a_*^2}}{2}}\right)Mc^2
\sim 1 \times 10^{54}\ {\rm erg} \left(\frac{M}{10 M_\odot}\right),
\end{eqnarray}
where the spin parameter is typically $a_*=a/M \sim 0.7$ after a merger
\citep[e.g.,][]{Zlochower_Lousto15}.

Post-merger spinning BHs should also exist in our Galaxy,
having a lot of energy in the spin.
The number of such BHs is estimated as
\begin{eqnarray}
N_{\rm BH} \sim \mathscr{R}_{\rm GW} n_{\rm gal}^{-1} H_0^{-1} 
\sim 7 \times 10^{4}\ {\rm galaxy}^{-1}
\left(\frac{\mathscr{R}_{\rm GW}}{70\,{\rm Gpc}^{-3}\,{\rm yr}^{-1}}\right),
\label{eq:NBH}
\end{eqnarray}
where we use $\mathscr{R}_{\rm GW} \sim 70$ Gpc$^{-3}$ yr$^{-1}$ \citep{LIGO_O1},
$n_{\rm gal} \sim 0.01$ Mpc$^{-3}$ is the number density of galaxies,
and $H_0^{-1} \sim 10^{10}$ yr is the Hubble time.
This estimate is applicable 
unless the merger rate changes very rapidly
in a time much shorter than the Hubble time.
Note that, although the large mass in GW150914 suggest 
a low-metallicity environment with $Z \lesssim Z_{\odot}/2$
\citep{LIGO_astro16},
our Galaxy had a low-metallicity environment in the past,
and also incorporated low-metallicity galaxies
in the hierarchical structure formation.
The total spin energy stored in the merged BHs in our Galaxy is
\begin{eqnarray}
E_{\rm tot}=N_{\rm BH} E_{\rm spin}
\sim 9 \times 10^{58}\ {\rm erg}
\sim 9 \times 10^{7} E_{\rm SN},
\label{eq:Etot}
\end{eqnarray}
where $E_{\rm SN} \sim 10^{51}$ erg is the kinetic energy of a supernova (SN).
This is comparable to the total energy of SNe that ever happened in our Galaxy,
i.e., $\sim 10^8$ SNe exploded during the Hubble time!
This is a robust lower limit on the total spin energy,
obtained by the GW observations for the first time.

A natural question arises: 
How much spin energy is extracted from the merged BHs in our Galaxy?
The spin energy of a BH can be extracted 
by a large-scale poloidal magnetic field threading the BH,
i.e., through Blandford-Znajek effect \citep{BZ77},
which is thought to produce a BH jet.
We show that a sufficient magnetic field is 
advected by the Bondi-Hoyle accretion from the interstellar medium (ISM)
and the jet power becomes comparable to the accretion rate,
which is larger than the radiative power of the accretion disk.
By taking into account the distributions of the ISM density, 
the BH mass and velocity,
we estimate the luminosity function and the total power of the BH jets.

Based on the estimate of the luminosities and the acceleration energy,
we suggest that the BH jets are potentially the origin of
high energy particles in our Galaxy.
There are enigmatic high-energy sources in our Galaxy,
such as still-unknown PeVatrons accelerating cosmic rays (CRs) 
up to the knee energy $\varepsilon_{\rm knee} \sim 3 \times 10^{15}$ eV and beyond,
sources of TeV CR positrons,
and unidentified TeV sources (TeV unIDs)
that are dominant in the very-high-energy gamma-ray sky.
These sources require only a small fraction of the spin energy $E_{\rm tot}$
and could be powered by the BH jets.

Our examination of the BH accretion and jet also suggests that 
it is very difficult to detect
an electromagnetic counterpart to a BH merger after a GW event.
In particular, the report of a GRB around the time of GW150914 
by the Fermi Gamma-ray Burst Monitor (GBM)
\citep{Connaughton+16}
is most likely irrelevant to the GW event.
This is consistent with a large number of follow-up searches after GW150914
\citep{Ackermann+16,Kasliwal+16,Troja+16,KamLAND16,Adrian-Martinez+16,Tavani+16,Abbott+16,Adriani+16,Evans+16,Palliyaguru+16,Auger+16,Abe+16,Morokuma+16,Evans+16b}.

The organization of this paper is as follows. 
In Section~\ref{sec:mechanism},
we discuss the physical mechanism of energy extraction from a spinning BH.
We find that the accretion disk typically results in 
the so-called magnetically arrested disk (MAD) state
and the magnetic field extracts the spin energy with the maximum efficiency
for producing a jet.
In Section~\ref{sec:LF},
we calculate the luminosity function of the BH jets
by considering the distributions of 
the BH mass, the peculiar velocity, the GW recoil velocity,
and the ISM density.
The luminosity function also gives the total power of the BH jets.
In Section~\ref{sec:obs}, we discuss the connections of the BH jets 
with high energy sources in our Galaxy,
such as PeVatrons, CR positron sources, and TeV unIDs.
In Section~\ref{sec:uncertain},
we encompass the uncertainties of our estimate on the total power
within a factor of $10^{\pm 3}$
by taking into account 
various effects such as the initial spin, the BH formation scenario, 
and the wind feedbacks.
This is much better than before; 
the factor was almost $10^{\pm \infty}$ before the GW detections.
In Section~\ref{sec:Fermi},
we show that BHs are difficult to keep accretion disks until the merger 
that are massive enough for making a detectable electromagnetic counterpart
for GW150914.
Section~\ref{sec:discuss} is devoted to the summary and discussions.
In Section~\ref{sec:history},
we clarify novel points of our work compared with previous studies.

\section{Extracting spin energy of GW150914-like Galactic BHs}
\label{sec:mechanism}

The spin energy of a BH can be extracted 
by a large-scale magnetic field threading the BH ergosphere.
The BH spin twists the magnetic field
and the twisted magnetic field carries energy outward as a Poynting jet.
This is the so-called BZ effect \citep{BZ77,Koide+02}.
Although the BH itself cannot keep the magnetic field 
because of the no-hair theorem,
accretion onto the BH can maintain the magnetic field on the BH.
In this section we consider a BH in the ISM
and estimate the luminosity of a BZ jet powered by the BH spin.
For typical parameters,
we find that 
the luminosity of a BH jet is comparable to the accretion rate
$L_{j} \approx {\dot M} c^2$,
with the accretion disk in the state of the so-called MAD.

\subsection{Bondi accretion from the ISM}\label{sec:Bondi}

The accretion rate onto a BH from the ISM is given by the Bondi-Hoyle rate
\citep{Hoyle_Lyttleton39,Bondi_Hoyle44,Bondi52},
\begin{eqnarray}
\dot{M}
&=& 4 \pi r_B^2 V \rho
=\frac{4\pi G^2 M^2 n \mu m_u}{V^{3}}
\nonumber\\
&\sim&
5 \times 10^{35} \ {\rm erg}\ {\rm s}^{-1}
\frac{1}{c^2}
\left(\frac{n}{10\,{\rm cm}^{-3}}\right)
\left(\frac{M}{10 M_\odot}\right)^2
\left(\frac{V}{10\,{\rm km}\,{\rm s}^{-1}}\right)^{-3}
\nonumber\\
&\sim& 4\times 10^{-4} \frac{{L}_{\rm Edd}}{c^2}
\left(\frac{n}{10\,{\rm cm}^{-3}}\right)
\left(\frac{M}{10 M_\odot}\right)
\left(\frac{V}{10\,{\rm km}\,{\rm s}^{-1}}\right)^{-3},
\label{eq:Mdot}
\end{eqnarray}
where $n$ is the number density of the ISM,
$m_u$ is the unified atomic mass unit,
the mean molecular weight is $\mu=1.41$ for the Milky Way abundance
and $\mu=2.82$ for molecular clouds \citep[e.g.,][]{Kauffmann+08},
$M$ is the mass of the merged BH,
$L_{\rm Edd}$ is the Eddington luminosity,
\begin{eqnarray}
r_B = \frac{GM}{V^2} 
\sim 1 \times 10^{15}\ {\rm cm}
\left(\frac{M}{10 M_\odot}\right)
\left(\frac{V}{10\,{\rm km}\,{\rm s}^{-1}}\right)^{-2}
\end{eqnarray}
is the Bondi radius, and
\begin{eqnarray}
V=\sqrt{c_s^2+v^2+v_{\rm GW}^2}
\end{eqnarray}
includes the (effective) sound speed $c_s$ of the ISM,
the center-of-mass velocity $v$ of the BH before the merger 
in the local ISM,
and the recoil velocity $v_{\rm GW}$ due to the GW emission at the merger.
The accretion rate ${\dot M}$ is proportional to $M^2 V^{-3} n$.
The discovery of a massive BH with mass $M \sim 60 M_{\odot}$ in GW150914
significantly increases the estimate of ${\dot M}$,
while the GW recoil tends to reduce it.
The ISM density spans many decades.
Thus we have to consider the distributions of mass, velocity, and density
to estimate the total power in Section~\ref{sec:LF}.

\subsection{Formation of an accretion disk and ADAF}\label{sec:disk}

The accreted matter forms an accretion disk for typical parameters
\citep{Fujita+98,Agol_Kamionkowski02}.
The ISM density has a turbulent fluctuation with a Kolmogorov spectrum
$\delta \rho/\rho \sim [L/(6 \times 10^{18}\,{\rm cm})]^{1/3}$
down to $\sim 10^{8}$ cm \citep{Armstrong+95,Draine11}.
As a BH travels in the ISM,
the accreting matter acquires a net specific angular momentum
\begin{eqnarray}
\ell \sim \frac{1}{4}\frac{\Delta \rho}{\rho} V r_B,
\end{eqnarray}
where $\Delta \rho/\rho=\delta \rho/\rho|_{L=2r_B}$
is the density difference across the accretion cylinder.\footnote{
A factor $1/4$ comes from the average over the accretion cylinder,
$\int_0^{r_B} r dr \int_0^{2\pi} d\theta (r \cos \theta)^2
/ r_B^2 \int_0^{r_B} r dr \int_0^{2\pi} d\theta = \frac{1}{4}$.
If a turbulent velocity dominates $V$,
the factor has a fluctuation
depending on the velocity direction.}
By equating this with the Keplerian angular momentum 
$\ell_K=\sqrt{GM r_{\rm disk}}$,
we obtain the radius of the resulting accretion disk,
\begin{eqnarray}
\frac{r_{\rm disk}}{r_S} &\sim&
\frac{1}{16}\left(\frac{2 GM/V^2}{6\times 10^{18}\,{\rm cm}}\right)^{2/3} 
\frac{c^2}{2V^2}
\nonumber\\
&\sim& 2\times 10^{5}
\left(\frac{M}{10 M_{\odot}}\right)^{2/3}
\left(\frac{V}{10\,{\rm km}\,{\rm s}^{-1}}\right)^{-10/3},
\label{eq:rdisk}
\end{eqnarray}
where $r_S=2GM/c^2$ is the Schwarzschild radius.
The disk radius could be decreased if the magnetic breaking is effective.

The accretion disk most likely forms hot, geometrically-thick accretion flow
such as advection-dominated accretion flow (ADAF)
\citep{Fujita+98}.
The accreted matter is heated and eventually ionized because
the collisional ionization rate is larger than
the accretion rate as well as the recombination rate
for typical parameters (see also Section~\ref{sec:Fermi}).
The accretion rate is much lower than the Eddington rate
as in Equation (\ref{eq:Mdot})
and hence the low density makes the cooling ineffective
\citep[][]{Ichimaru77,Narayan_Yi94,Narayan_Yi95,Kato+08,Yuan_Narayan14}.
The radiated energy from ADAF is much less than the total generated energy
and almost all energy is advected into the BH
(see also Section~\ref{sec:wind}).
For example, the luminosity of bremsstrahlung emission from electrons
is only 
$L_{\rm brem}\sim (\alpha_{\rm QED}/\alpha^2)(m_e/m_u) 
({\dot M}c^2/L_{\rm Edd}) {\dot M} c^2
\ll {\dot M} c^2$,
where $\alpha_{\rm QED}$ is the fine-structure constant,
$\alpha$ is the viscous parameter,
and $m_e$ is the electron mass.
As shown below, this is much smaller than the jet luminosity.
Thus we concentrate on the jet in this paper
and consider the disk emission in the future papers
\citep{Matsumoto+16}.

A transition to a cold standard disk outside the hot disk is not expected
for typical parameters,
although this is common in BH X-ray binaries \citep[e.g.,][]{Esin+97,Kato+08}.
The reason is that at the initial radius in Equation~(\ref{eq:rdisk}),
the disk is already hot (ionized) and
the maximum accretion rate of the ADAF solution \citep{Abramowicz+95} 
is larger than the accretion rate in Equation (\ref{eq:Mdot}), 
i.e., cooling is ineffective.
Then we do not also expect soft X-ray transients 
(or X-ray novae) caused by the accumulation of the accreted matter 
at some radius
because the thermal instability 
due to recombination is absent for the ionized flow \citep[e.g.,][]{Kato+08}.

\subsection{Blandford-Znajek jet from a MAD state}\label{sec:BZ}

\begin{figure}
  \begin{center}
    \includegraphics[width=\columnwidth]{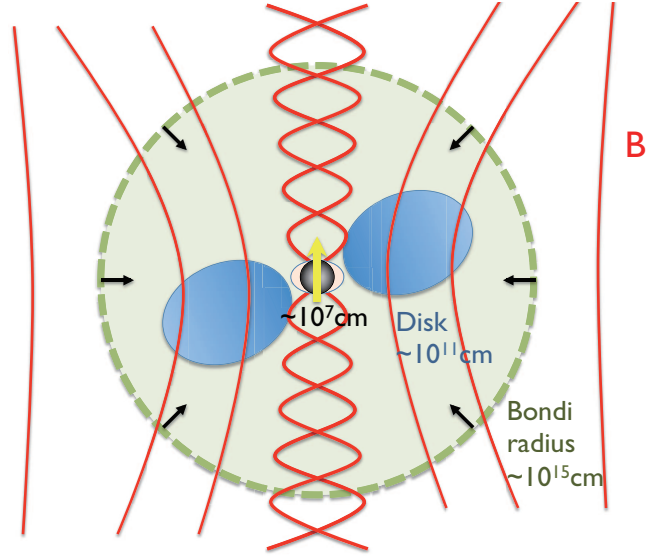}
  \end{center}
  \caption{
    Schematic picture of a Blandford-Znajek jet from a spinning BH 
    that is accreting from the ISM.
  }
  \label{fig:BZ}
\end{figure}

The accretion of the ISM also drags magnetic fields into the BH
(see Figure~\ref{fig:BZ}).
The magnetic fields are well frozen in the accreting fluid
because the loss time of the magnetic flux in the ISM is much longer 
than the accretion time \citep{Nakano+02}.
The formed disk is also thick, being able to 
advect the magnetic flux inward \citep{Lubow+94,Cao11}.
The coherent length of the magnetic field in the ISM is 
much larger than the Bondi radius,
approximately about the scale of energy injection by SNe and stellar winds
$\sim 1$--$10$ pc \citep{Han+04}.
Then the magnetic flux conservation implies 
the magnetic field strength on the horizon
\begin{eqnarray}
B_H \sim \left(\frac{r_B}{r_H}\right)^2 B_{\rm ISM},
\label{eq:flux_cons}
\end{eqnarray}
where $B_{\rm ISM}$ is the magnetic field strength in the ISM,
and $r_H=\frac{1}{2}\left(1+\sqrt{1-a_*^2}\right) r_S$
is the radius of the BH horizon.
On the other hand, for a given accretion rate, 
there is a maximum strength of the magnetic field on the horizon,
\begin{eqnarray}
B_H &\sim & \left.\sqrt{\frac{4GM{\dot M}}{r^3 v_r}}\right|_{r=r_H}
\nonumber\\
&\sim & 4 \times 10^{7}\ {\rm G}
\left(\frac{n}{10\,{\rm cm}^{-3}}\right)^{1/2}
\left(\frac{V}{10\,{\rm km}\,{\rm s}^{-1}}\right)^{-3/2},
\label{eq:B_MAD}
\end{eqnarray}
because the pressure of the magnetic field,
\begin{eqnarray}
p_B=\frac{B^2}{8\pi},
\end{eqnarray}
can not exceed the ram pressure of the accreting matter,
\begin{eqnarray}
p_{a} = \frac{GM\Sigma}{r^2}
\sim \frac{GM {\dot M}}{2\pi r^3 v_r},
\end{eqnarray}
where $\Sigma={\dot M}/2\pi r v_r$ is the surface density,
$v_r \equiv \epsilon v_{\rm ff}$ is the radial velocity,
$v_{\rm ff}=\sqrt{3GM/4\pi r}$ is the free-fall time,
and $\epsilon \sim 0.05$ is suggested
by the numerical simulations and observations 
\citep{Tchekhovskoy+11,Zamaninasab+14}.
Although accumulation of the magnetic flux with the same polarity 
makes a magnetic barrier \citep{BR76},
the accretion continues through a magnetic flux
via magnetic interchange instability \citep[e.g.,][]{Arons_Lea76,McKinney+12}.
Such a magnetically-dominated state is the so-called MAD
\citep{BR76,Narayan+03,Tchekhovskoy+11}.
The MAD state is realized 
if $B_H$ in Equation (\ref{eq:flux_cons}) is larger than that in Equation (\ref{eq:B_MAD}), i.e.,
\begin{eqnarray}
B_{\rm ISM} &>& B_{\rm crit} 
\equiv \left(\frac{r_H}{r_B}\right)^2
\left.\sqrt{\frac{4GM{\dot M}}{r^3 v_r}}\right|_{r=r_H}
\nonumber\\
&\sim & 1 \times 10^{-10}\,{\rm G}
\left(\frac{V}{10\,{\rm km}\,{\rm s}^{-1}}\right)^{5/2}
\left(\frac{n}{10\,{\rm cm}^{-3}}\right)^{1/2}.
\end{eqnarray}
This is usually satisfied for typical $B_{\rm ISM} \sim 3 \mu$G.
Thus the formed disk is likely MAD.

A spinning BH immersed in large-scale poloidal magnetic fields
releases energy through the BZ effect
with a Poynting luminosity
\begin{eqnarray}
L_{j} \approx \frac{\kappa}{4\pi c} \Omega_H^2 \Psi_{\rm BH}^2,
\label{eq:LBZ}
\end{eqnarray}
where $\kappa \approx 0.05$ \citep{Tchekhovskoy+11}, 
$\Omega_H=a_*c/2r_H$ is the angular frequency of the BH,
and $\Psi_{\rm BH} \sim \pi r_H^2 B_H$ is a magnetic flux on the BH.
For a MAD state $p_B \sim p_a$, the BZ luminosity
is calculated as
\begin{eqnarray}
L_{j} \sim \left(\frac{\kappa}{\epsilon}
\sqrt{\frac{\pi^3}{12}\frac{r_S}{2r_H}}\right)
a_*^2 {\dot M} c^2
\approx {\dot M} c^2
\label{eq:LBZ=Mdot}
\end{eqnarray}
for a typical spin parameter after the merger $a_*\approx 0.7$.
In the following we will use $L_{j} \approx {\dot M} c^2$
to estimate the jet luminosity of the merged BHs.\footnote{
We have confirmed that the results are almost similar even if we use
$L_{j} \approx a_*^2 {\dot M} c^2$.
}

Note that the net angular momentum direction of the accretion flow
changes on a timescale of crossing the density fluctuation
$t_a \sim r_B/V \sim 40\,{\rm yr}
(M/10M_{\odot}) (V/10\,{\rm km}\,{\rm s}^{-1})^{-3}$.
However the angular momentum vector near the BH
is forced to align with the BH spin
by the Bardeen-Petterson effect \citep{Bardeen_Petterson75}.
In addition, although the direction of the poloidal magnetic fields
is generally different from the BH spin direction,
the magneto-spin alignment is also realized by
the frame-dragging effect \citep{McKinney+13}.
Therefore we can consider that 
the direction of the jet is the same as that of the BH spin.

\section{Luminosity function of GW150914-like Galactic BH jets}\label{sec:LF}

Since the accretion rate depends on $n M^2 V^{-3}$ that spans many decades,
we calculate the luminosity function of jets from GW150914-like merged BHs 
in our Galaxy as
\begin{eqnarray}
\frac{dN}{d{\dot{M}}}
&=&
N_{\rm BH} \int dm_1\, \frac{dp(m_1)}{dm_1}
\int dm_2\, \frac{dp(m_2|m_1)}{dm_2}
\int dv\, \frac{df(v)}{dv}
\nonumber\\
&\times&  
\int dn\, \frac{d\xi(n)}{dn}
h(m_1,m_2,v)
\delta\left[\dot{M}(n, m_1, m_2, v)-\dot{M}\right],
\label{eq:LF1}
\end{eqnarray}
where $dp(m_1)/dm_1$ and $dp(m_2|m_1)/dm_2$ are the distributions of BH masses
(Section~\ref{sec:mass}),
$df(v)/dv$ is the distribution of the pre-merger velocity
(Section~\ref{sec:velocity}),
$d\xi(n)/dn$ is the distribution of the ISM density 
(Section~\ref{sec:density}),
and $h(m_1,m_2,v)$ is the correction factor
due to the scale heights of the ISM phases and BH distributions
(Section~\ref{sec:height}).
First, the delta function can be integrated over $v$ analytically as
\begin{eqnarray}
\frac{dN}{d{\dot M}}
&=&
N_{\rm BH} \int dm_1\, \frac{dp(m_1)}{dm_1}
\int dm_2\, \frac{dp(m_2|m_1)}{dm_2}
\int dn\, \frac{d\xi(n)}{dn}
\nonumber\\
&\times&  
h(m_1,m_2,v_0)
\frac{df(v_0)}{dv}
\frac{V_{v=v_0}^2}{3 v_0 {\dot M}},
\end{eqnarray}
where $v_0^2\equiv (4\pi G^2 M^2 n \mu m_u/{\dot M})^{2/3}-c_s^2-v_{\rm GW}^2$
should be positive, otherwise the integrant vanishes.
The other integrals are computed numerically.
We adopt $N_{\rm BH} \sim 7 \times 10^4$ BHs galaxy$^{-1}$ as a fiducial value,
corresponding to the GW event rate
$\mathscr{R}_{\rm GW} \sim 70$ Gpc$^{-3}$ yr$^{-1}$ \citep{LIGO_O1}
in Equation~(\ref{eq:NBH}).

\subsection{Mass function}\label{sec:mass}

We assume a Salpeter-like mass function for the primary BH,
\begin{eqnarray}
\frac{dp(m_1)}{dm_1} = C m_1^{-\gamma},
\end{eqnarray}
with a uniform distribution of the secondary mass,
\begin{eqnarray}
\frac{dp(m_2|m_1)}{dm_2} = \frac{1}{m_1-M_{\min}},
\end{eqnarray}
where $\gamma=2.35$,
$M_{\min} \le m_2 \le m_1 \le M_{\max}$,
$M_{\min}=5M_{\odot}$,
$M_{\max} = 50 M_\odot$, and
$C=(\gamma-1)/(M_{\min}^{1-\gamma}-M_{\max}^{1-\gamma})$.
Such mass functions are inferred by the observations of massive stars
\citep{Sana+12,Kobulnicky+14}.
Similar mass functions\footnote{
LIGO imposes $m_1 + m_2 < 100M_{\odot}$ instead of $M_{\max} = 50 M_\odot$.
} are adopted 
by the analysis of LIGO O1 data \citep{LIGO_O1},
and are consistent with the GW observations.
Note that the total luminosity is dominated by heavy masses for $\gamma<3$.
In this respect, GW150914 is crucial by raising the maximum mass $M_{\max}$ 
and hence the expected luminosity more than was previously thought
(cf. $M_{\max}=13 M_{\odot}$ was adopted in \citet{Agol_Kamionkowski02}.
Note also $\gamma=0.35$ in \citet{Agol_Kamionkowski02}).

\subsection{Velocity distribution before a merger}\label{sec:velocity}

The velocity distribution for GW150914-like BHs before mergers is described by 
a Maxwell-Boltzmann distribution,
\begin{eqnarray}
\frac{df(v)}{dv}=\sqrt{\frac{2}{\pi}} \frac{v^2}{\sigma_v^3} 
\exp \left(-\frac{v^2}{2 \sigma_v^2}\right),
\label{eq:velocity}
\end{eqnarray}
where an isotropic Gaussian approximation 
is enough for our order-of-magnitude estimates.
As a fiducial value, we take the velocity dispersion $\sigma_v=40$ km s$^{-1}$
by considering the isolated binary formation scenario.
From a theoretical point of view, massive star progenitors are born
from molecular clouds and their velocity dispersion is initially low
$\sigma_v \sim 10$ km s$^{-1}$ \citep{Binney_Merrifield98}.
Unless the BH formation is associated with an exceptionally large kick
due to such as asymmetric mass ejection,
the resulting BHs have also low velocities.
If the kick velocity is inversely proportional to the mass 
following the momentum conservation,
the kick velocity of neutron stars implies
$\sigma_v \sim 200$ km s$^{-1} (1.4 M_\odot/M)
\sim 30$ km s$^{-1} (M/10M_{\odot})^{-1}$.
Older stars tend to have larger velocity dispersion and
$\sigma_v=40$ km s$^{-1}$ is reasonable 
for progenitors with metallicity $Z \lesssim 0.5 Z_\odot$
\citep{Binney_Merrifield98}.
From an observational point of view,
the rms distance $\sim 410$ pc from the Galactic plane 
for BH low-mass X-ray binaries,
corresponding to a scale height of $290$ pc,
suggest a velocity dispersion of $\sigma_v \sim 40$ km s$^{-1}$
\citep{White_vanParadijs96}.
Although there are exceptions such as
GRO 1655-40 with a peculiar velocity $v \sim -114$ km s$^{-1}$ 
\citep{Brandt+95,Mirabel+02}
and XTE J1118+480 with $v \sim 145$ km s$^{-1}$ \citep{Mirabel+01},
two populations likely exist with low and high kick velocities,
similarly to neutron stars \citep{Cordes_Chernoff98,Pfahl+02}.
On the other hand, these observations are not for high-mass systems.
In addition, these estimates are subject to systematic errors in the distance
\citep{Repetto+12}.
The most reliable estimate is based on the astrometric observations 
\citep{Miller-Jones14}.
Although there is only one sample for a high-mass system,
the BH high-mass X-ray binary Cygnus X-1 has a relatively low proper motion
$\sim 20$ km s$^{-1}$
\citep{Chevalier_Ilovaisky98,Mirabel_Rodrigues03,Reid+11}.

We discuss a high-velocity case $\sigma_v=200$ km s$^{-1}$ 
later in Section~\ref{sec:scenario}.

\subsection{ISM Density}\label{sec:density}

\begin{table*}[t]
  \centering
  \caption{Five ISM phases.
    The density distribution from $n_1$ to $n_2$ with an index $\beta$, 
    the volume filling fraction $\xi_0$,
    the (effective) sound velocity $c_s$ 
    (including the turbulent velocity for cold phases),
    and the scale height of the disk $H_d$ are shown.
  }
  \label{tab:ISM}
  \begin{tabular}{lllllll} \hline
    Phase & $n_1$ [cm$^{-3}$] & $n_2$ [cm$^{-3}$] & $\beta$ & $\xi_0$ & $c_s$ [km s$^{-1}$] & $H_d$  \\ \hline \hline
    Molecular clouds & $10^2$ & $10^5$ & $2.8$ & $10^{-3}$ & 10 & 75 pc \\
    Cold ${\rm H}_{\rm I}$ & $10$ & $10^2$ & $3.8$ & $0.04$ & 10 & 150 pc \\
    Warm ${\rm H}_{\rm I}$ & 0.3 & $-$ & $-$ & 0.35 & 10 & 0.5 kpc \\
    Warm ${\rm H}_{\rm II}$ & 0.15 & $-$ & $-$ & 0.2 & 10 & 1 kpc \\
    Hot ${\rm H}_{\rm II}$ & 0.002 & $-$ & $-$ & 0.4 & 150 & 3 kpc \\ \hline
  \end{tabular}
\end{table*}

We consider five phases of the ISM as listed in Table~\ref{tab:ISM};
the molecular clouds consisting mostly of H$_2$,
the cold neutral medium consisting of H$_{\rm I}$ clouds (cold H$_{\rm I}$),
the warm neutral medium in thermally equilibrium 
with cold H$_{\rm I}$ (warm H$_{\rm I}$),
the warm ionized medium (warm H$_{\rm II}$),
and the hot ionized medium (hot H$_{\rm II}$)
\citep{Bland-Hawthorn_Reynolds00,Ferriere01,Heyer_Dame15,Inutsuka+15}.
For each phase, we use the probability distribution of the number density,
\begin{eqnarray}
\frac{d\xi(n)}{dn} = D \xi_0 n^{-\beta},\quad (n_1<n<n_2)
\label{eq:density}
\end{eqnarray}
where $n_1$, $n_2$ and $\beta$ are given in Table~\ref{tab:ISM}
\citep{Berkhuijsen99},
$D=(\beta-1)/(n_1^{1-\beta}-n_2^{1-\beta})$,
and $\xi_0 = \int \frac{d\xi}{dn} dn$ is the volume filling fraction
\citep{Scoville_Sanders87,Clemens+88,Agol_Kamionkowski02}.\footnote{
For the cases without $n_2$ in Table~\ref{tab:ISM}, we use a delta function.}
Each phase has its scale height $H_d$ in the Galactic disk.
We assume that the hot H$_{\rm II}$ phase has a sound velocity $c_s=150$ km s$^{-1}$ 
corresponding to a temperature $T \sim 10^{6}$ K,
while the other phases have effective $c_s \sim 10$ km s$^{-1}$ 
(corresponding to $T \sim 2 \times 10^{4}$ K)
because these phases (even with $T< 2 \times 10^{4}$ K)
have also a turbulent velocity $\sim 10$ km s$^{-1}$
in approximately pressure balance with each other.
The parameters in Table~\ref{tab:ISM} are similar to those in
Agol \& Kamionkowski (2002) \citep{Agol_Kamionkowski02}.

\subsection{Scale height}\label{sec:height}

The BHs have their scale height $H(v_z)$ in the Galactic disk.
Each phase of the ISM has also its own scale height $H_d$ 
in Table~\ref{tab:ISM}.
Then the number of BHs in each phase is corrected by a factor
\begin{eqnarray}
h(m_1,m_2,v)=\min\left[1,\frac{H_d}{H(v_z)}\right].
\end{eqnarray}
For simplicity, we make a one-dimensional analysis of the vertical structure,
neglecting the coupling of the vertical and horizontal motions.
The scale height $H\left(v_z\right)$
is determined by the velocity in the $z$-direction,
\begin{eqnarray}
\frac{1}{2}v_z^2=\Phi_z\left[H(v_z)\right],
\end{eqnarray}
where the $z$-component of the velocity is $v_z^2=\frac{1}{3}(v^2+v_{\rm GW}^2)$
and the gravitational potential in the $z$-direction,
\begin{eqnarray}
\frac{\Phi_z(z)}{2\pi G}=K \left(\sqrt{z^2+Z^2}-Z\right)+Fz^2,
\end{eqnarray}
where $Z \sim 180$ pc, $K=48 M_\odot$ pc$^{-2}$, and $F=0.01 M_\odot$ pc$^{-3}$
\citep{Kuijken_Gilmore89a,Kuijken_Gilmore89b}.
This simple model is sufficient for our order-of-magnitude estimates.

\subsection{GW recoil velocity}

Merged BHs receive a recoil due to the anisotropic GW emission
\citep{Bonnor_Rotenberg61,Peres62,Bekenstein73}.
GWs carry linear momentum if two merging BHs have different masses and/or finite spins.
Fitting formulas for the recoil velocity are obtained
by using numerical simulations in the post-Newtonian-inspired forms
\citep{Zlochower_Lousto15,Campanelli+07,Baker+07}.
The recoil velocity for a merger of non-spinning BHs is well approximated by
\begin{eqnarray}
v_{\rm GW} &=& A \eta^2 \sqrt{1-4\eta} \left(1+B \eta\right),
\end{eqnarray}
where 
$A=1.20 \times 10^{4}$ km s$^{-1}$, $B=-0.93$ \citep{Gonzalez+07,Fitchett83},
and $\eta={m_1 m_2}/{(m_1+m_2)^2}$ is the symmetric mass ratio.
The maximum value is $v_{\rm GW} \sim 175$ km s$^{-1}$ for $\eta=0.195$.
GW150914 has $\eta \sim 0.247$ and hence $v_{\rm GW} \sim 61$ km s$^{-1}$,
and GW151226 has $\eta \sim 0.226$ and hence $v_{\rm GW} \sim 150$ km s$^{-1}$.
Although we have to extrapolate the formula to small $\eta$,
this does not affect our result so much.

We do not consider the spin-induced recoil
because we are now assuming that the pre-merger spin is low 
to make a conservative estimate on the released energy from BHs.
If the pre-merger spin is high,
the pre-merger BHs, which are much more abundant than the merged BHs,
can release energy through the BZ mechanism
even before the merger without the GW recoil.
This case will be discussed in Section~\ref{sec:spin}.
Current observations of GWs show 
that the primary BH has a spin of $<0.7$ at 90\% confidence
with no evidence for spins being both large and strongly aligned.
For GW151226, the effective spin parameter is $0.21^{+0.20}_{-0.10}$
and may imply spinning BHs \citep{LIGO_O1}.
For reference, if the spin parameters parallel to the orbital axis are 
$a_{*2\parallel} \sim 0.2$ and $a_{*1\parallel} \sim 0$ 
with the same masses $\eta \sim 1/4$,
the recoil velocity is about
$v_{\rm GW} \sim 40$ km s$^{-1}$,
while it is $v_{\rm GW} \sim 260$ km s$^{-1}$ 
if the in-plane spins are $a_{*2\perp} \sim 0.2$ and $a_{*1\perp} \sim 0$ 
with the same masses $\eta \sim 1/4$ \citep{Zlochower_Lousto15}.

\subsection{Results of the luminosity function}

\begin{figure}
  \begin{center}
    \includegraphics[width=\columnwidth]{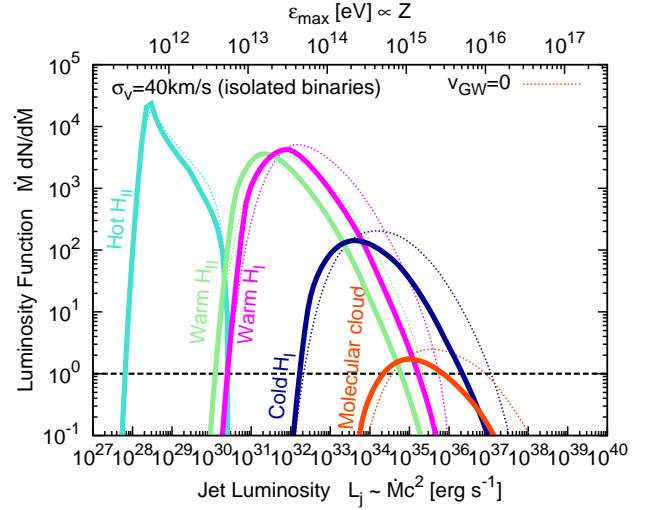}
  \end{center}
  \caption{
    Luminosity function of the BH jets 
    from GW150914-like merged BHs in our Galaxy
    for the isolated binary formation scenario 
    with the velocity dispersion $\sigma_v=40$ km s$^{-1}$,
    calculated from Equation~(\ref{eq:LF1}).
    Each line corresponds the ISM phase where the BHs reside 
    (see Table~\ref{tab:ISM}).
    We can see that the most luminous source in our Galaxy has 
    $L_{j} \sim 10^{36}$ erg s$^{-1}$ for this scenario.
    Dotted lines are calculated by setting the GW recoil velocity $v_{\rm GW}=0$.
    The upper horizontal axis is the maximum energy 
    of accelerated particles allowed by the Hillas condition
    for a given luminosity with the charge of accelerated particles $Z=1$ 
    in Equation~(\ref{eq:emax}).
  }
  \label{fig:LF1}
\end{figure}

Figure~\ref{fig:LF1} shows the luminosity function of
the BH jets from accreting BHs in our Galaxy
for the fiducial case (see Table~\ref{tab:model} for the other cases),
calculated from Equation~(\ref{eq:LF1}).
Each line corresponds the ISM phase where the BHs reside.
As the accretion rate is proportional to the ISM density
in Equation~(\ref{eq:Mdot}),
the jet luminosity is brighter for BHs in denser ISM such as molecular clouds.
On the other hand, brighter jets are rarer
because the volume filling fraction
of denser medium is smaller in the ISM
as in Table~\ref{tab:ISM}.
We can find that the brightest sources in our Galaxy
(with the number $dN/d\,{\rm log}\,{\dot M} \sim 1$)
have $L_{j} \sim 10^{36}$ erg s$^{-1}$
mostly residing in the cold ${\rm H_I}$,
while fainter sources are more abundant.

The GW recoil effect reduces the luminosity by an order-of-magnitude
as we can see from the dotted lines, 
which are calculated by setting $v_{\rm GW}=0$.
This reduction is approximately determined 
by the $V^{-3}$ dependence of the accretion rate in Equation~(\ref{eq:Mdot}) as
$\sim (v_{\rm GW}/\sigma_v)^{-3}
\sim (100\,{\rm km}\,{\rm s}^{-1}/40\,{\rm km}\,{\rm s}^{-1})^{-3}
\sim 0.06$.
Note also that the luminosity function for the hot $H_{\rm II}$ phase has a peak
because this phase has a large $c_s$ and so $V \sim c_s$ has little dispersion.

\begin{figure}
  \begin{center}
    \includegraphics[width=\columnwidth]{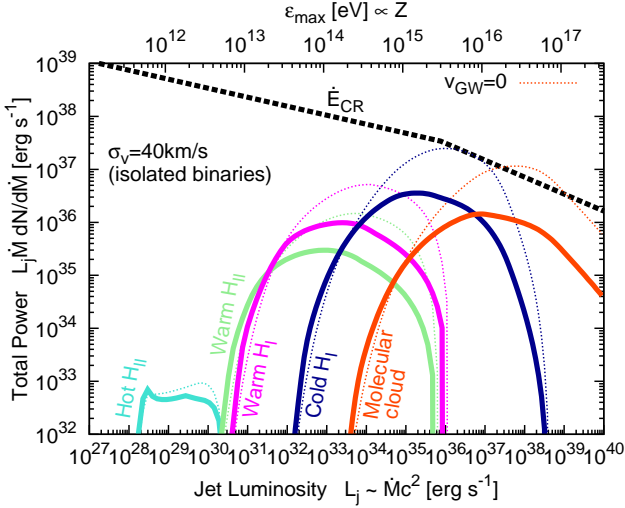}
  \end{center}
  \caption{
    Same as Figure~\ref{fig:LF1} but with the vertical axis 
    multiplied by $L_{j} \sim {\dot M}c^2$.
    We can find that the most energy is generated by BHs 
    in the cold H$_{\rm I}$ medium.
    Dashed line plots the required power spectrum for supplying the observed CRs.
    We can see that the total power is comparable to 
    that of CRs above the knee energy at around $\sim 3$ PeV
    within the model uncertainties that are listed in Table~\ref{tab:model}.
  }
  \label{fig:power}
\end{figure}

Figure~\ref{fig:power} is the same as Figure~\ref{fig:LF1} but 
with the vertical axis multiplied by $L_{j} \sim {\dot M}c^2$.
This makes it clear that the most energy is generated 
by BHs in the cold H$_{\rm I}$ medium.
We can read the total power,
\begin{eqnarray}
P_{\rm tot}=\int L_{j} \frac{dN}{d{\dot M}} d{\dot M}
\sim 10^{37}\ {\rm erg}\ {\rm s}^{-1}
\left(\frac{N_{\rm BH}}{7 \times 10^{4}}\right),
\label{eq:budget}
\end{eqnarray}
which is very roughly derived by
$P_{\rm tot} \sim N_{BH} \times \xi_0^{{\rm H}_{\rm I}} 
\times {\dot M}(10\,{\rm cm}^{-3}, 5 M_{\odot}, 5 M_{\odot}, 100\,{\rm km}\,{\rm s}^{-1}) c^2
\times (M_{\max}/M_{\min})^{3-\gamma}
\sim 7 \times 10^{4} \times 0.04 \times 5 \times 10^{32}\,{\rm erg}\,{\rm s}^{-1} 
\times 4.5
\sim 0.6 \times 10^{37}\,{\rm erg}\,{\rm s}^{-1}$.
Note that the velocity dependences of $\dot M$ and $f(v)$ cancel with each other
and the low velocity BHs have smaller scale height than the cold H$_{\rm I}$.
The total power is approximately $\sim 3 \times 10^{-5}$ of that of SN explosions
${E_{\rm SN}}/{100\,{\rm yr}}
\sim 3 \times 10^{41}\ {\rm erg}\ {\rm s}^{-1}$.
This is small but comparable to the required power
for some high-energy particles in our Galaxy.
Based on these results, we will discuss observational implications in the next section.

\section{Observational implications}\label{sec:obs}

\subsection{PeVatrons}\label{sec:PeVatron}

Particle acceleration is ubiquitous in the BH jet system
as manifested in active galactic nuclei (AGNs),
gamma-ray bursts (GRBs) and X-ray binaries \citep{Longair11}.
The maximum acceleration energy is limited by the source size, i.e.,
the so-called Hillas condition,
$\varepsilon_{\max}=Z q B r/\Gamma$,
where $Z$ is the charge of accelerated particles,
$\Gamma$ is the Lorentz factor of the acceleration region,
and $B$ is the lab-frame magnetic field.
This can be written in terms of the Poynting luminosity
$L_{j} \sim 2 \pi \theta_j^2 r^{2} c (B^2/8\pi)$,
where the magnetic field carries an energy density $B^2/8\pi$ at a radius $r$
with the jet opening angle $\theta_j$ \citep{Norman+95,Blandford00,Waxman04}.
With the causality condition $\Gamma \theta_j \gtrsim 1$, we have
\begin{eqnarray}
\varepsilon_{\max}=Zq B \frac{r}{\Gamma}\sim \frac{2Zq}{\Gamma \theta_j} \sqrt{\frac{L_{j}}{c}}
\lesssim 3.5\ {\rm PeV}\ Z \left(\frac{L_{j}}{10^{36}\,{\rm erg}\,{\rm s}^{-1}}\right)^{1/2}.
\label{eq:emax}
\end{eqnarray}
Therefore bright sources, i.e., massive BHs in dense ISM,
are potential ``PeVatrons'' accelerating particles beyond PeV energy
\citep{Barkov+12,Kotera_Silk16}.
We plot $\varepsilon_{\max}$ on the upper horizontal axis 
in Figures~\ref{fig:LF1} and \ref{fig:power}.
Possible acceleration sites are discussed in Section~\ref{sec:discuss}.

In Figure~\ref{fig:power},
we also plot the required power spectrum for supplying the observed CRs
by using
$L_0 (\varepsilon/\varepsilon_{\min}^{\rm CR})^{2-s} \frac{d\varepsilon}{\varepsilon}
=L_0 (\varepsilon/\varepsilon_{\min}^{\rm CR})^{2-s} \frac{d{\dot M}}{2{\dot M}}$
where the spectral index is $s=2.34$ below the knee \citep{Genolini+15}\footnote{
The intrinsic spectral index of CRs $s$ is unobservable and
different from the observed one by the diffusion coefficient index,
which is usually obtained from observations of the boron-to-carbon ratio
\citep{Evoli+15,Oliva+15,Genolini+15}.
}
and $0.3$ higher above the knee \citep{Blumer+09,Gaisser+13}.
SN remnants are commonly believed to supply
most Galactic CRs from 
the peak energy $\varepsilon_{\min}^{\rm SN}=1$ GeV
to the knee $\varepsilon_{\max}^{\rm SN}=3$ PeV.
The normalization $L_0$ is determined by the fact
that a fraction $\epsilon_{\rm CR}=0.1$ of the SN kinetic energy
can yield CRs,
\begin{eqnarray}
\frac{\epsilon_{\rm CR} E_{\rm SN}}{100\,{\rm yr}}
=\int_{\varepsilon_{\min}^{\rm SN}}^{\varepsilon_{\max}^{\rm SN}} L_0 
\left(\frac{\varepsilon}{\varepsilon_{\min}^{\rm SN}}\right)^{2-s} 
\frac{d\varepsilon}{\varepsilon}.
\end{eqnarray}

From Figure~\ref{fig:power} we can see that 
the BH jets can produce comparable energy to that required for
the observed CRs at and beyond the knee energies $\gtrsim 3$ PeV,
taking the model uncertainties into account 
(see Section~\ref{sec:uncertain}).
The origin of these CRs is not known \citep[e.g.,][]{Hillas05,Blumer+09,Gaisser+13}.
Currently known gamma-ray sources, including even SN remnants,
do not show the characteristic PeVatron spectrum extending 
without a cutoff or break to tens of TeV \citep{Aharonian13},
with a possible exception of the Galactic center Sagittarius A* \citep{HESS16}.
Even if the SN remnants are responsible for CRs up to the knee,
the transition from Galactic to extragalactic CRs
occurs between the knee and the ankle.
Ultra-high-energy CRs above the ankle are extragalactic
because of the observed isotropy \citep{Abreu+10,Abbasi+16}.\footnote{
There could be possible hot spots \citep{Abbasi+14,Aab+15}.}
If the knee corresponds to the proton cutoff
and the source composition is solar,
the rigidity-dependent cutoffs extending beyond the knee
are not sufficient to fill the observed all-particle flux 
\citep{Hillas05}.
This implies a second (Galactic) component at energies between the knee and the ankle, sometimes called ``component B''.
Our results suggest that the BH jets might be PeVatrons
and/or fill the gap between the knee and the ankle.
An unnatural point of this possibility is that
the BH jets are totally irrelevant to the SN remnants.
It is just a coincidence that the CR fluxes from two kinds of sources
are the same within a factor.
There are also orders-of-magnitude uncertainties 
in the estimate of the total power of the BH jets 
(see Section~\ref{sec:uncertain}).
Furthermore it is difficult to calculate the CR spectrum and
the acceleration efficiency at present.
Nevertheless the BH jets can potentially accelerate 
the CRs at and beyond PeV energy
with the flux comparable to the observations.

\subsection{Cosmic-ray positrons and electrons}\label{sec:positron}

The CR positron fraction (the ratio of positrons to electrons plus positrons)
has been measured by the PAMELA satellite \citep{Adriani+09}
and more precisely by the AMS-02 experiment \citep{Aguilar+13}.
The observed positron fraction rises from $\sim 10$ GeV at least to $\sim 300$ GeV,
indicating the presence of nearby positron sources within $\sim 1$ kpc.
Although the dark matter annihilation or decay scenario is now severely constrained by
other messengers,
there are still many astrophysical candidates and the ture origin is unclear
\citep[e.g.,][]{Serpico12,Ioka10,Kashiyama+11,Fujita+09,Kohri+16}.

The BH jets could accelerate electrons and positrons preferentially
if the jets are not contaminated by baryons \citep{Barkov+12}
but associated with the pair cascade
(see also Section~\ref{sec:discuss}).
The maximum energy of particle acceleration is enough for producing
the observed positrons as in Equation~(\ref{eq:emax}).
The required total power for the positron excess is
about $\sim 10^{-4}$ of that of SN explosions, i.e., about $\sim 3 \times 10^{37}$ erg s$^{-1}$.
This is comparable to that of the BH jets in Equation~(\ref{eq:budget})
within the model uncertainties
(see Section~\ref{sec:uncertain}).
Therefore the BH jets are eligible to join the possible sources,
although it is again difficult to estimate the spectrum and efficiency of 
the positron acceleration.

A BH jet likely forms an extended nebula in the ISM, 
similarly to an AGN cocoon/lobe \citep{Begelman+89b} and a GRB cocoon
\citep{Ramirez-Ruiz+02,Mizuta_Ioka13}.
A BH jet collides with the ISM at the jet head.
The shocked matter goes sideways forming a cocoon.
Although the cocoon pressure initially collimates the jet,
the collimation radius expands and finally reaches 
the termination (reverse) shock.
The maximum size of the termination shock is given by the condition that
the jet pressure balances with the ram pressure of the ISM,
$L_{j}/2 \pi \theta^2 r_h^2 c \sim n \mu m_u V^2$,
i.e.,
\begin{eqnarray}
r_h &\sim & \sqrt{\frac{L_{j}}{2\pi \theta^2 c n \mu m_u V^2}}
\sim 2\ {\rm pc}
\left(\frac{L_{j}}{10^{36}\,{\rm erg}\,{\rm s}^{-1}}\right)^{1/2}
\nonumber\\
&\times&
\left(\frac{n}{10\,{\rm cm}^{-3}}\right)^{-1/2}
\left(\frac{V}{10\,{\rm km}\,{\rm s}^{-1}}\right)^{-1}
\left(\frac{\theta}{0.1}\right)^{-1},
\label{eq:rhead}
\end{eqnarray}
at which point the jet is completely bent by the ISM
and dissipated into the cocoon.
The cocoon is also extended along the direction of the proper motion,
leading to a more or less spherical shape.
The forward shock of the BH jet nebula expands with 
a velocity $v_c \sim (L_{j}/n \mu m_u)^{1/5} t^{-2/5}$
and a size $r_c \sim v_c t$,
slowing down to $v_c \sim V$ at the maximum size
$r_{c,\max} \sim (L_{j}/n m_u V^3)^{1/2}
\sim 80$ pc $(L_{j}/10^{36}\,{\rm erg}\,{\rm s}^{-1})^{1/2}
(n/10\,{\rm cm}^{-3})^{-1/2}
(V/10\,{\rm km}\,{\rm s}^{-1})^{-3/2}$.
The BH jet nebula is similar to an old pulsar wind nebula (PWN).
Then the CR electrons and positrons likely escape from the nebula to the ISM 
without adiabatic cooling \citep{Kashiyama+11}.
Radiative cooling such as synchrotron emission
limits the maximum energy of electrons and positrons,
which depends on the propagation time and the magnetic field in the nebula 
\citep{Kawanaka+10}.
Future observations beyond TeV energies by CALET, DAMPE and CTA will 
probe such leptonic PeVatrons \citep{Kobayashi+04,Kawanaka+11}.

\subsection{Unidentified TeV gamma-ray sources}\label{sec:unID}

The Galactic Plane survey carried out by HESS led to the discovery of
dozens of TeV gamma-ray sources.
Among these, the most abundant category is dark accelerators, so-called 
TeV unidentified sources (TeV unIDs), 
which have no clear counterpart at other wavelengths
\citep{Aharonian+05,Aharonian+06,Aharonian+08}.
They lie close to the Galactic plane, suggesting Galactic sources.
Their power-law spectra with an index of 2.1--2.5 imply
a connection with CR accelerators.
They are extended $\Delta \Theta \sim 0.05$--$0.3^{\circ}$,
corresponding to a physical size of
$\sim 3\,{\rm pc}\,(\Delta \Theta/0.2^{\circ}) (D/{\rm kpc})$ 
for an unknown distance $D$.
Still, their unID nature prevents us to identify their origin
\citep[e.g.,][]{Yamazaki+06,deJager+09,Ioka_Meszaros10}.

\begin{figure}
  \begin{center}
    \includegraphics[width=\columnwidth]{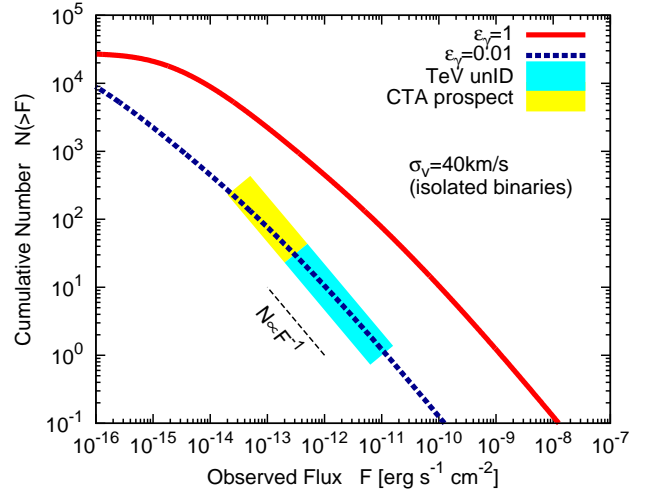}
  \end{center}
  \caption{
    Flux distribution expressed by the cumulative number of sources
    above a given flux, i.e., log $N(>F)$--log $F$ plot,
    with Equations~(\ref{eq:dN/dF}) and (\ref{eq:LF1}).
    This is compared with the observations of TeV unIDs.
    Both are comparable if the gamma-ray efficiency is 
    $\epsilon_\gamma \sim 10^{-2}$.
    CTA could detect additional $\sim 300$ BH jets in the near future.
  }
  \label{fig:flux}
\end{figure}

In Figure~\ref{fig:flux}, we plot the observed flux distribution of the TeV unIDs 
at energies $\varepsilon_\gamma >0.2$ TeV
in terms of the cumulative number of sources above a flux $N(>F)$,
i.e., log $N$--log $F$ plot.
In order to compare it with the BH jets,
we calculate the flux distribution from the luminosity function
in Equation~(\ref{eq:LF1})
by integrating the number of sources above a given (bolometric) flux 
$F = L_{j}/4\pi D^2 \sim {\dot M} c^2/4\pi D^2$
as
\begin{eqnarray}
\frac{dN}{dF}=\int \frac{dN}{d{\dot M}}
\frac{4\pi D^2}{c^2}
\frac{DdD d\theta}{\pi R_d^2},
\label{eq:dN/dF}
\end{eqnarray}
where we approximate the spatial distribution of the BH jets by
a thin uniform disk with a radius $R_d=15$ kpc
and the distance of the Sun to the Galactic center $R_\odot=8$ kpc.
A thin approximation is applicable if the observed distance is
larger than the scale height $\sim 300$ pc 
for the fiducial case (see Table~\ref{tab:model}).

Figure~\ref{fig:flux} shows that
the flux distribution is comparable with that of TeV unIDs
if the gamma-ray efficiency is about $\epsilon_\gamma \sim 10^{-2}$
for the fiducial parameters (see Table~\ref{tab:model}).
Note that the IC cooling time of 10 TeV electrons is $\sim 10^5$ yr. 
If the age is $\sim 10^5$ yr, the TeV gamma-ray flux is 
$\sim 0.1$--$0.02\, L_e$,
implying that $\epsilon_e \sim 0.1$--$0.5$.
This is comparable with values considered in GRB jets and PWNe.
If this is the case,
the CTA (Cherenkov Telescope Array) observatory
will increase the number of TeV unIDs up to $\sim 300$ 
by improving the sensitivity by about an order of magnitude
in the near future \citep{CTA13}.
Note that the flux distribution follows $N(>F) \propto D^{2} \propto F^{-1}$
if the spatial distribution is disk-like,
which is different from $N(>F) \propto F^{-1.5}$ for the 3D Euclidian space.
The uniform disk approximation is acceptable
for the current observations, which have not reached the Galactic center yet.
For future observations, we have to consider 
the high density region near the Galactic center.

The nebular size in Equation~(\ref{eq:rhead}) is also consistent with the 
extended nature of TeV unIDs.
The BH jet nebula also evades
strong upper limits in X-rays with a TeV to X-ray flux ratio up to $\gtrsim 50$
\citep{Matsumoto+07,Bamba+07,Matsumoto+08,Bamba+09,Fujinaga+11,Sakai+11}.
This is because the physical parameters such as the energy density
and the magnetic field are similar to those of an old PWN.
Their emission spectra have the unID nature
thanks to the old age \citep{Yamazaki+06,deJager+09,Ioka_Meszaros10}.
In addition, the ADAF disk is radiatively inefficient.
The X-ray flux of the ADAF disk is about
$F_{\nu} \sim
(\alpha_{\rm QED}/\alpha^2) ({m_e}/{m_u})
({{\dot M}c^2}/{L_{\rm Edd}})
{\dot M}/4\pi D^2 m_e
\sim 1 \times 10^{-18}\,{\rm erg}\,{\rm s}^{-1}\,{\rm cm}^{-2}\,{\rm keV}^{-1}$
$\left(\alpha/0.1\right)^{-2}$
$\left({n}/{10\,{\rm cm}^{-3}}\right)^2$
$\left({M}/{10 M_\odot}\right)^3$
$\left({V}/{10\,{\rm km}\,{\rm s}^{-1}}\right)^{-6}$
$\left({D}/{{\rm kpc}}\right)^{-2}$,
below the current limit.

\section{Model uncertainties}\label{sec:uncertain}

Although the GW observations significantly narrow down 
the possible parameter space,
in particular putting a lower bound on the number of spinning BHs
in Equation~(\ref{eq:NBH}),
there are still large uncertainties about the model parameters
and the estimate for the BH jet power.
In this section, we clarify the range of the uncertainties by considering
four representative effects: 
the initial BH spin (Section~\ref{sec:spin}),
the velocity distribution depending on the binary BH formation scenario 
(Section~\ref{sec:scenario}),
the accretion rate profile changed by the disk wind (Section~\ref{sec:wind}),
and the feedback on the ISM by the outflow (Section~\ref{sec:feedback}).
These effects on the model parameters and 
the resulting total power are summarized 
in Table~\ref{tab:model}.
We enclose the uncertainty of the total power for the BH jets
within a factor of $10^{\pm 3}$,
which is much better than before.

\begin{table*}
  \centering
  \caption{Possible uncertainties of the model parameters and 
    the resulting total power $P_{\rm tot}$ are summarized.
    The isolated binary formation scenario is the fiducial case.
    The column with ``$-$'' equals the fiducial value.
    The model parameters are the number of BH jets $N_{\rm BH}$,
    the dispersion of the velocity distribution $\sigma_v$ in Equation~(\ref{eq:velocity}), 
    the initial BH spin $a_*^i$, 
    the accretion rate profile $s$ in Equation~(\ref{eq:profile}), 
    and the duty cycle $\mathscr{D}$ reduced by the feedback 
    on the ISM by the disk outflow.
  }
  \label{tab:model}
  \begin{tabular}{lllllll} \hline
    & Number & Velocity & Spin & Disk & Duty cycle & Total power \\ 
    & $N_{\rm BH}$ & $\sigma_v$ [km s$^{-1}$] & $a_*^i$ & $s$ & $\mathscr{D}$ & $P_{\rm tot}$ [erg s$^{-1}$] \\ \hline \hline
    Isolated binary (fiducial) & $7 \times 10^4$ & $40$ & $0$ & $0$ & $1$ & $\sim 10^{37}$ \\
    High spin & $10^{8}$ & $-$ & $0.2$ & $-$ & $-$ & $\sim 10^{37} \times 10^{3}$ \\
    Stellar cluster & $-$ & $200$ & $-$ & $-$ & $-$ & $\sim 10^{37} \times 10^{-2}$ \\
    Wind & $-$ & $-$ & $-$ & $1$ & $-$ & $\sim 10^{37} \times 10^{-2}$ \\
    Feedback & $-$ & $-$ & $-$ & $1$ & $10^{-1}$ & $\sim 10^{37} \times 10^{-3}$ \\ \hline
  \end{tabular}
\end{table*}

\subsection{Initial spin}\label{sec:spin}

If BHs have spins before the mergers,
the BHs can launch BZ jets without the mergers.
Such spinning BHs could result from the massive stellar collapse.
The total number of BHs in our Galaxy is 
about $N_{\rm BH} \sim 10^{8}$ \citep{Shapiro_Teukolsky83},
$\sim 10^{3}$ times larger than that of the merged BHs in Equation~(\ref{eq:NBH}).\footnote{
The number would be $N_{\rm BH} \sim 10^{10}$ 
if dark matter were composed of primordial BHs that are relevant to GWs,
while it is suggested that the fraction of primordial BHs in dark matter
is small $\sim 10^{-4}$
by observations of GWs and CMB spectral distortion \citep{Sasaki+16}.
}
In addition the GW recoil is absent without a merger,
increasing the total power by a factor of ten 
as shown in Figure~\ref{fig:power}.
Then the total power is larger than the fiducial value
by a factor of $\sim 10^{4} (a_*^i/0.7)^2$ altogether,
that is, $P_{\rm tot} \sim 10^{41}$ erg $(a_*^i/0.7)^2$,
where the $(a_*^{i})^2$ dependence comes from 
that of the BZ luminosity in Equation~(\ref{eq:LBZ}).

GW observations show no evidence for large spins.
Probably the initial spin would be small for most BHs
because the massive star progenitors with solar metallicity
lose the angular momentum by stellar wind \citep{Heger+03,Hirschi+05}.
Because of the same reason, the BH mass is also smaller than the fiducial case
\citep{LIGO_astro16}, reducing the total power of the BH jets.
In low metallicity, the wind is weak and the resulting BH spin may be high
\citep{Yoon_Langer05,Hirschi+05,Kinugawa+16c}.
A rapid rotation of the progenitors could lead to
a chemically homogeneous evolution without a common envelope phase,
avoiding a merger before the BH formation
\citep{Mandel_deMink16,Marchant+16}.
However the number of such BHs is much less than $N_{\rm BH} \sim 10^{8}$.
The event rate of GRBs, which likely produce spinning BHs,
is comparable to that of the BH mergers.
Although some BHs in X-ray binaries might have high spins,
these measurements are subject to systematic errors 
\citep{Remillard_McClintock06}.
For GW151226, the effective spin parameter is $0.21^{+0.20}_{-0.10}$ 
\citep{LIGO_O1}.
So we tentatively take $a_*^i \sim 0.2$ as an upper limit 
in Table~\ref{tab:model}.
This is the most extreme case because the total power 
is comparable to that of SN explosions,
${E_{\rm SN}}/{100\,{\rm yr}} \sim 3 \times 10^{41}\ {\rm erg}\ {\rm s}^{-1}$.

\subsection{Binary BH formation scenario}\label{sec:scenario}

The accretion rate and the resulting jet luminosity sensitively depend on 
the velocity of the BH in Equation~(\ref{eq:Mdot}).
We have adopted $\sigma_v=40$ km s$^{-1}$ as a fiducial value 
for the isolated binary formation scenario
\citep{Tutukov_Yungelson93,Dominik+15,Belczynski+16,Lipunov+16}
in Equation~(\ref{eq:velocity}).

The GW150914 masses favor low metallicity below $0.5 Z_{\odot}$ 
\citep{LIGO_astro16}.
The extreme case is zero metallicity Population III stars
\citep{Kinugawa+14,Kinugawa+16a,Kinugawa+16b,Kinugawa+16c}.
If BHs form in very low metallicity $< 0.01 Z_{\odot}$,
the GW events may be dominated by recent BH mergers 
in dwarf galaxies \citep{Lamberts+16}
because the low metallicity allows a small initial separation of a BH binary.
Then the merged, spinning BHs are incorporated into our Galaxy 
relatively recently,
joining in the halo component with a velocity dispersion of
$\sigma_v \sim 200$ km s$^{-1}$.

Another scenario is 
the dynamical binary formation in a dense stellar cluster
\citep{Kulkarni+93,Sigurdsson_Hernquist93,PortegiesZwart_McMillan00,Rodriguez+15,Rodriguez+16a,Mapelli16}.
In a high-density stellar environment, 
BHs dynamically interact and form binaries.
Since the interaction is frequent in the clusters,
most of the BH mergers may occur outside the clusters 
following dynamical ejection.
The escape velocity of the clusters is smaller than that of our Galaxy.
Thus the merged BHs are floating in our halo
with a velocity dispersion of $\sigma_v \sim 200$ km s$^{-1}$.

Primordial BHs are also a possible candidate
\citep[e.g.,][]{Bird+16,Sasaki+16,Nakamura+97,Ioka+98,Ioka+99},
although this scenario requires a fine tuning 
in the primordial density fluctuation.
In this case, the BHs reside in our halo with $\sigma_v \sim 200$ km s$^{-1}$.

\begin{figure}
  \begin{center}
    \includegraphics[width=\columnwidth]{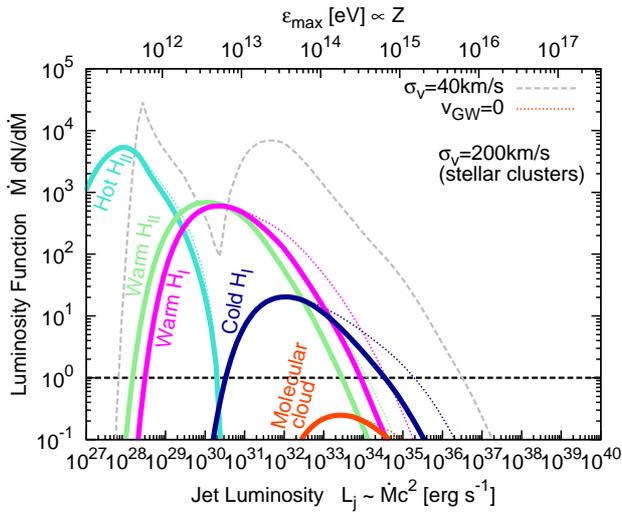}
  \end{center}
  \caption{
    Same as the fiducial case in Figure~\ref{fig:LF1} 
    except for the dispersion of the velocity distribution 
    $\sigma_v=200$ km s$^{-1}$.
    The total luminosity function for the fiducial case 
    $\sigma_v=40$ km s$^{-1}$ is also plotted by
    a gray dashed line.
  }
  \label{fig:LF2}
\end{figure}

Figure~\ref{fig:LF2} shows the case of $\sigma_v \sim 200$ km s$^{-1}$.
Compared with the fiducial case $\sigma_v \sim 40$ km s$^{-1}$
(gray dashed line),
the luminosity and hence the total power 
are reduced by a factor of $\sim 10^{2}$.
This factor is roughly given by the velocity dependence of the accretion rate,
$\sim (40\,{\rm km}\,{\rm s}^{-1}/200\,{\rm km}\,{\rm s}^{-1})^3 \sim 0.008$.
The GW recoil effect becomes less significant than the fiducial case
because the velocity dispersion is comparable with the recoil velocity.

\subsection{Wind}\label{sec:wind}

It remains highly uncertain how much of the accreting matter 
at the Bondi radius reaches the BH \citep{Yuan_Narayan14}.
Some supermassive BH systems with jets
seem to require the Bondi accretion rates calculated from the observed
gas temperature and density to power the observed jets
\citep{Allen+06,Rafferty+06,Russell+13}.
On the other hand, 
the ADAF model implies positive Bernoulli parameter for the inflow
in the self-similar regime,
which suggests that hot accretion flows could have outflows
\citep{Narayan_Yi94,Narayan_Yi95}.
The mass outflows make the accretion profile decrease inward
approximately in a power-law form,
${\dot M}(r)={\dot M}(r_{\rm disk}) \left({r}/{r_{\rm disk}}\right)^{s}$,
as in the adiabatic inflow-outflow solutions (ADIOS) model
\citep{Blandford_Begelman99,Blandford_Begelman04}.
The index is limited to $0 \le s < 1$ by the mass and energy conservation,
but has not been determined yet \citep{Yuan_Narayan14}.
The least accretion case corresponds to $s \approx 1$.
Recent 3D general relativistic MHD simulations
suggest that $s \approx 1$ continues down to $20 r_S$,
below which the mass flux is constant $s=0$ \citep{Yuan+15}.
This is also implied by an analytical study \citep{Begelman12}.
If we adopt this least accretion case,
the accretion rate of the BH is given by
\begin{eqnarray}
{\dot M}_{\rm BH} \approx {\dot M} \left(\frac{20 r_S}{r_{\rm disk}}\right)^{s},
\quad {\rm if}\ r_{\rm disk}>20r_S,
\label{eq:profile}
\end{eqnarray}
with $s \approx 1$ where the disk radius $r_{\rm disk}$ 
is given by Equation~(\ref{eq:rdisk}).
Correspondingly, the luminosity of the BH jet is reduced by the same factor
$\left({20 r_S}/{r_{\rm disk}}\right)^{s}$.

\begin{figure}
  \begin{center}
    \includegraphics[width=\columnwidth]{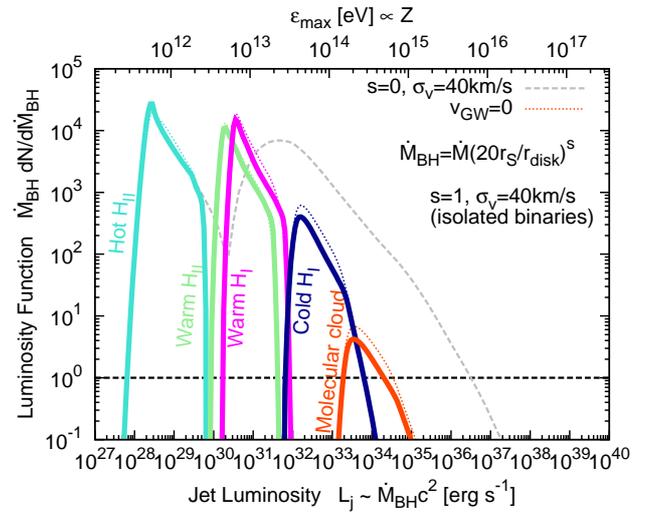}
  \end{center}
  \caption{
    Same as the fiducial case in Figure~\ref{fig:LF1} 
    except for the accretion rate of the BH
    ${\dot M}_{\rm BH}={\dot M} (20 r_S/r_{\rm disk})^{s}$
    with $s=1$, which is reduced by the wind.
    The total luminosity function for the fiducial case 
    $s=0$ is also plotted by a gray dashed line.
  }
  \label{fig:LF3}
\end{figure}

Figure~\ref{fig:LF3} shows the luminosity function
using the accretion rate of a BH in Equation~(\ref{eq:profile}).
Compared with the fiducial case $s=0$ (gray dashed line),
the luminosity and the total power 
are reduced by a factor of hundred.
This factor is roughly given by the ratio
$r_{\rm disk}/20 r_S$ in Equation~(\ref{eq:rdisk}).
The GW recoil effect becomes less significant than the fiducial case
because the disk radius $r_{\rm disk}$ and the accretion rate ${\dot M}$ 
have similar dependences on the velocity.

\subsection{Feedback}\label{sec:feedback}

Feedback from radiation, jets and winds on the surrounding ISM 
could be crucial for estimating the total power of the BH jets,
as frequently argued in the context of supermassive BHs
\citep{Yuan_Narayan14}.
In the Galactic BH case, 
the radiative feedback is weak because the ADAF disk is much fainter than
the Eddington luminosity.
The radiation may ionize the ISM around the Bondi radius,
but once ionized, the cross section 
for the interaction between the ISM and photons
decreases by many orders of magnitude, reducing the radiative feedback.
The jet feedback is also not strong
because, although the jet dominates the energy output,
its penetration ability makes the dissipation scale large
as shown in Equation~(\ref{eq:rhead}).
A large amount of ISM is capable of radiating the injected energy.

The most influential feedback would be due to the wind from the disk,
if it exists.
If the wind is efficient with $s \approx 1$ in Equation~(\ref{eq:profile}),
even a small efficiency of the wind feedback 
$\epsilon_w \gtrsim 10^{-6} (M/10 M_{\odot})^{2/3} 
(V/10\,{\rm km}\,{\rm s}^{-1})^{-4/3}$
is able to heat the ISM at the Bondi radius to blow away,
$\epsilon_w {\dot M}_{\rm BH} c^2 > {\dot M} V^2$.
A larger efficiency $\epsilon_w\sim 0.03$--$0.001$ 
is implied by simulations \citep{Sadowski+16,Yuan+15,Ohsuga_Mineshige11}.
On the other hand, the wind will stop
if the mass accretion at the Bondi radius is terminated.
Therefore we expect that the BH activity is intermittent
with some duty cycle $\mathscr{D}$ if the wind feedback exists.

A rough estimate of the duty cycle is as follows.
The wind is somewhat collimated initially when it is released from the disk.
If it were spherical, the wind would not be launched because
the ram pressure of the Bondi accretion onto the disk exceeds
that of the wind.
The $4\pi$ solid angle of the ISM is affected by the wind
after the wind is decelerated by the ISM, 
which will happen outside the Bondi radius
because the ram pressure of the accretion is a decreasing function of the radius
and the wind goes straight inside the Bondi radius.
Thus the accretion continues at least 
for the dynamical time at the Bondi radius,
\begin{eqnarray}
t_a = \frac{r_B}{V} \sim 40\ {\rm yr} 
\left(\frac{M}{10 M_{\odot}}\right) 
\left(\frac{V}{10\,{\rm km}\,{\rm s}^{-1}}\right)^{-3}.
\end{eqnarray}
The injected energy during the active time is about
\begin{eqnarray}
E_{a}&=&\epsilon_w {\dot M}_{\rm BH} c^2 t_a
\sim 3 \times 10^{40}\ {\rm erg}
\left(\frac{\epsilon_w}{3\%}\right)
\left(\frac{n}{10\,{\rm cm}^{-3}}\right)
\nonumber\\
&\times&\left(\frac{M}{10 M_{\odot}}\right)^{11/3}
\left(\frac{V}{10\,{\rm km}\,{\rm s}^{-1}}\right)^{-8/3}.
\end{eqnarray}
The injected energy produces a wind remnant
and the lifetime of the remnant is about 
$t_{\rm merge} \sim  t_{\rm PDS} \times 153 (E_{51}^{1/14} n_0^{1/7}/V_6)^{10/7}
\sim 2\times 10^{6}$ yr $E_{51}^{3/14} n_0^{-4/7} (E_{51}^{1/14} n_0^{1/7}/V_6)^{10/7}$
according to the notation in \citet{Cioffi+88},
which gives
\begin{eqnarray}
t_{\rm merge} &\sim& 300\ {\rm yr}
\left(\frac{E_a}{10^{40}\,{\rm erg}}\right)^{31/98} 
\left(\frac{n}{10\,{\rm cm}^{-3}}\right)^{-18/49}
\nonumber\\
&\times& \left(\frac{V}{10\,{\rm km}\,{\rm s}^{-1}}\right)^{-10/7}.
\end{eqnarray}
Therefore the duty cycle is roughly 
\begin{eqnarray}
\mathscr{D} &\sim& \frac{t_a}{t_{\rm merge}} \sim 0.1
\left(\frac{V}{10\,{\rm km}\,{\rm s}^{-1}}\right)^{-0.73}
\nonumber\\
&\times& \left(\frac{n}{10\,{\rm cm}^{-3}}\right)^{-0.051}
\left(\frac{M}{10\,M_{\odot}}\right)^{-0.16}.
\end{eqnarray}
We use $\mathscr{D} \sim 10^{-1}$ in Table~\ref{tab:model}.

\section{On Fermi GBM events associated with GW150914}\label{sec:Fermi}

The Gamma-ray Burst Monitor (GBM) on board the Fermi satellite
reported a 1 sec-lasting weak GRB 0.4 seconds after GW150914.
Assuming the redshift of GW150914, $z=0.09^{+0.03}_{-0.04}$,
the luminosity in 1 keV--10 MeV is $1.8^{+1.5}_{-1.0} \times 10^{49}$ erg s$^{-1}$
\citep{Connaughton+16}.
This was unexpected 
and prompted many theoretical speculations
\citep{Loeb16,Zhang16,Perna+16,Li+16,Veres+16,Cardoso+16,Kimura+16}.
The anti-coincidence shield (ACS) of the Spectrometer on board INTEGRAL (SPI)
put upper limits on the gamma-ray emission 
with similar fluxes \citep{Savchenko+16}.
The GBM result also depend on the analysis of low count statistics
\citep{Greiner+16}.
No counterpart is observed for GW151226 and LVT151012
\citep{Racusin+16}.
Future follow-ups would be finally necessary 
to confirm or defeat the GBM detection
\citep{Yamazaki+16,Morsony+16,Murase+16}.

If the signal were caused by the merged BH, 
the BH would be surrounded by matter.
The size of the matter distribution is
$r_{m} \sim 1 \times 10^{8}$ cm
so that the accretion time is
\begin{eqnarray}
t_{\rm acc}&=&\frac{1}{\alpha \Omega_K} \left(\frac{r_{m}}{H}\right)^2
\sim 1.4\ {\rm sec}
\left(\frac{\alpha}{0.1}\right)^{-1}
\left(\frac{M}{60 M_{\odot}}\right)^{-1/2}
\nonumber\\
&\times&\left(\frac{r_m}{1 \times 10^{8}\,{\rm cm}}\right)^{3/2}
\left(\frac{H/r_m}{0.3}\right)^{-2},
\end{eqnarray}
where $\alpha$ is the viscosity parameter,
$H$ is the disk scale height,
and $\Omega_K=\sqrt{GM/r_m^3}$ is the Kepler rotation frequency.
The mass of the matter should be larger than
$M_{m} \gtrsim 10^{-5} \theta_j^2 M_{\odot}$
where $\theta_j$ is the opening angle of the GRB jet.

The accretion from the ISM affects the matter surrounding a BH.
In particular, it can evaporate a possible dead disk 
which were invoked for the GBM event
\citep{Perna+16,Kimura+16}.
A dead disk is assumed to be cold and neutral due to the small mass,
suppressing the magnetorotational instability and hence the viscosity,
and remains unaccreted, keeping matter for producing the gamma-ray event.
However the accretion from the ISM forms
a hot disk sandwiching the dead disk and heating its surface.\footnote{
Note that the accretion does not stop even onto a binary
because the disk is thick.
}
The surface temperature develops a gradient,
being greater than $T \gtrsim 10^4$ K 
(corresponding to the sound velocity 
$v_i \sim \sqrt{k_B T/m_u} \sim 10$ km s$^{-1}$)
for the ionized atmosphere.
The density $n_i$ at the base of the ionized atmosphere
is determined by the pressure balance
$n_i v_i^{2} \sim n(r) v_h^2$
where $n(r) \sim n (r_B/r)^{3/2}$ and $v_h \sim \sqrt{GM/r}$
are the density and the thermal velocity of the hot disk.
Given $n_i$ and $v_i$,
we can estimate the mass evaporation rate 
\citep[cf.][]{Hollenbach+94} from the dead disk
as
\begin{eqnarray}
{\dot M}_{\rm eva}
&\sim& 2\pi r^2 n_i v_i m_u
\sim 1\times 10^{15}\,{\rm g}\,{\rm s}^{-1}
\left(\frac{r}{10^{12}\,{\rm cm}}\right)^{-5/2}
\nonumber\\
&\times& \left(\frac{M}{60 M_{\odot}}\right)^{4}
\left(\frac{n}{1\,{\rm cm}^{-3}}\right)^{5/2}
\left(\frac{V}{40\,{\rm km}\,{\rm s}^{-1}}\right)^{-9/2}.
\end{eqnarray}
Then the evaporation time of the dead disk with mass $M_{m}$ is
\begin{eqnarray}
t_{\rm eva} &\sim& 10^{6}\,{\rm yr}
\left(\frac{M_{m}}{10^{-5} M_{\odot}}\right)
\left(\frac{r}{10^{12}\,{\rm cm}}\right)^{5/2}
\left(\frac{M}{60 M_{\odot}}\right)^{-4}
\nonumber\\
&\times& \left(\frac{n}{1\,{\rm cm}^{-3}}\right)^{-5/2}
\left(\frac{V}{40\,{\rm km}\,{\rm s}^{-1}}\right)^{9/2},
\end{eqnarray}
which is shorter than $\sim 10^{10}$ yr, 
the merger time of the BH binary with a separation $r \sim 10^{12}$ cm,
for the fiducial case.
One should keep in mind that the above equation is rather sensitive to 
parameters $M$, $n$ and $V$. 
For example, BH binaries could have a dead disk if they are formed
in a low-density environment. 
However, for typical parameters, the merged BH would not have a dead disk, 
implying that the GBM event is not related with GW 150914 in the dead disk scenario.

We can also make a second argument
that a time-reversal of this event seems to encounter physical difficulty.
Let's go back in time, say $t_b \sim 1000$ sec before the merger.
Still the two BHs should be surrounded by the matter.
The size of the matter distribution should be larger
$r_{m} \sim 10^{10}\,{\rm cm}\, (t_b/10^{3}\,{\rm sec})^{2/3}
(\alpha/0.1)^{2/3}
(M/60 M_{\odot})^{1/3}
(H/r_m/0.3)^{4/3}$,
otherwise the matter is swallowed by the BHs before the merger.
The bounding energy of this matter is only 
a fraction of the rest mass energy of the matter,
\begin{eqnarray}
\frac{G M M_{m}/r_{m}}{M_m c^2}
&\sim& 10^{-3} 
\left(\frac{t_b}{10^{3}\,{\rm sec}}\right)^{-2/3}
\left(\frac{\alpha}{0.1}\right)^{-2/3}
\nonumber\\
&\times& \left(\frac{M}{60 M_{\odot}}\right)^{2/3}
\left(\frac{H/r_m}{0.3}\right)^{-4/3}.
\end{eqnarray}
This ratio is much smaller than the wind efficiency 
$\epsilon_w \sim 0.1$ of a super-Eddington accretion disk,
so that such matter is easily blown away by the disk wind.
As long as a possible dead disk is ionized by the ISM accretion 
(that occurs unless we consider low $n$ and high $V$), 
we have encountered an unlikely setup. 
Note that a fraction of the matter $M_{\rm m}$ should accrete onto the BHs 
before the merger,
otherwise a fine-tuning is needed
because the time $t_b$ is much larger than the event duration $t_{\rm acc}$.
The accretion is super-Eddington, 
even if only a fraction of the matter accretes,
and should be accompanied by a strong disk wind
as suggested by numerical simulations \citep{Ohsuga+05,Jiang+14,Sadowski+14}.
Therefore it is difficult to keep the matter near the BH before the merger
and the BH mergers unlikely accompany 
observable prompt electromagnetic counterparts.


\section{Summary and Discussions}\label{sec:discuss}

We suggest possible connections between 
the BH mergers observed by GWs
and the high energy sources of TeV-PeV particles in our Galaxy.
The GW observations give a lower limit on the number of 
merged and hence highly-spinning BHs as 
$\sim 70000 (\mathscr{R}_{\rm GW}/70\,{\rm Gpc}^{-3}\,{\rm yr}^{-1})$,
and the spinning BHs produce relativistic jets
by accreting matter and magnetic fields from the ISM.
We calculate the luminosity function, the total power, and 
the maximum acceleration energy of the BH jets,
and find that 
the BH jets are eligible for PeVatrons, sources of CR positrons, and TeV unIDs.
The BH jets form extended nebulae like PWNe.
If they are observed as TeV unIDs,
additional $\sim 300$ nebulae will be discovered by CTA.

We quantify the uncertainties of the estimate for the total power of the BH jets
within a factor of $10^{\pm 3}$,
which is much better than before the GW detections,
by considering the initial BH spin,
the velocity distribution depending on the formation scenario,
the accretion profile changed by the wind,
and the feedback by the outflow (Table~\ref{tab:model}).
The uncertainties will be reduced by the GW observations,
in particular, of the BH spins.
It is also important to clarify
the feedback by the wind from the sub-Eddington accretion disk
on the Bondi-Hoyle accretion.

Our considerations on the BH accretion and jet
imply that the electromagnetic counterparts to BH mergers
including the Fermi GBM event after GW150914 are difficult to detect
with the current sensitivity.
The accretion from the ISM can evaporate
the cold neutral dead disk around the BH.
A slight accretion before the merger can also
blow away the surrounding matter if any.
These should be considered as constraints on dead disk models 
for prompt electromagnetic counterparts of the BH-BH merger.

Although we do not go into detail in this paper,
there are several sites of particle acceleration for a BH jet.
First, the BH magnetosphere acts as a particle accelerator like pulsars
if a gap arises with an electric field along the magnetic field
\citep{Hirotani+16,Hirotani_Pu16}.
The gamma-ray emission associated with leptonic acceleration
may be detectable for nearby sources 
although its luminosity is usually much smaller than the BZ luminosity.
Second, the internal shocks in the jet are possible like GRBs and AGNs.
As long as $B \propto \Gamma/r$ during the propagation,
the maximum acceleration energy is the same as Equation~(\ref{eq:emax}).
Third, the jet dissipates the magnetic energy 
when the MHD approximation breaks down.
This happens when the plasma density 
drops below the Goldreich-Julian density
\citep{Goldreich_Julian69},
which is the minimum density required for shielding the electric field.
The comoving plasma density is given by
$n'_p \sim L/4\pi r^2 m_u c^3 \Gamma^2 (1+\sigma)$
where $L \sigma/(1+\sigma)$ is the BZ luminosity in Equation~(\ref{eq:LBZ}),
$\sigma$ is the ratio of the Poynting to particle energy flux,
$\Gamma$ is the Lorentz factor of the jet,
and we should make an appropriate correction if jets are leptonic.
The comoving Goldreich-Julian density beyond the light cylinder 
$r_{\ell}=c/\Omega_H = 2r_H/a_*$ is 
$n'_{\rm GJ} \sim (\Omega_H/2\pi q c) (r_H/r_{\ell})^3 (r_{\ell}/r \Gamma)$.
By equating $n'_p$ with $n'_{\rm GJ}$, we obtain
the radius at which the MHD breaks down,
\begin{eqnarray}
r_{\rm MHD} &\sim& \sqrt{\frac{\pi \kappa L}{\sigma (1+\sigma) c}}
\frac{q}{m_u c^2}
\frac{r_H}{a_*^2 \Gamma}
\sim 2 \times 10^{13}\,{\rm cm}
\left[\sigma (1+\sigma)\right]^{-1/2}
\nonumber\\
&\times& a_{*}^{-2} \Gamma^{-1}
\left(\frac{L}{10^{35}\,{\rm erg}\,{\rm s}}\right)^{1/2}
\left(\frac{M}{10 M_{\odot}}\right).
\end{eqnarray}
Forth, the termination (reverse) shock of the jet 
at the radius in Equation~(\ref{eq:rhead})
is also a plausible site
like a hot spot of AGNs and a pulsar wind nebula for pulsars.
The jet could be subject to instability,
injecting energy into a cocoon/lobe before reaching the termination shock.
The shocks between the cocoon and the ISM are also possible sites
of particle acceleration.
Note that the BH Cygnus X-1 is surrounded by a ring-like structure in radio,
which may be formed by the interaction 
between a jet/cocoon and the ISM \citep{Gallo+05}.

We do not discuss the disk emission in detail.
Nearby BH disks with bremsstrahlung, synchrotron, and inverse Compton emission
could be detected in the future surveys \citep{Matsumoto+16}.
The accretion disks could also accelerate nonthermal particles
and contribute to the observed cosmic rays \citep{Teraki+16}.
An on-axis BH jet may be also observable 
if the beaming factor is larger than $\sim 0.01$.
These are interesting future problems.

\section*{Acknowledgements}

We thank Takashi Nakamura, Tsvi Piran, and Masaru Shibata for helpful comments.
This work is supported by
KAKENHI 24103006, 24000004, 26247042, 26287051 (K.I.).




\bibliographystyle{mnras}
\bibliography{ref} 




\appendix

\section{Comparison with previous works}\label{sec:history}

Studying the accreting BHs in our Galaxy 
has a long history from 1920's Eddiongton era.
However the observations of GWs set 
a lower limit on the number of spinning BHs for the first time.
Our paper gives the first considerations on the Galactic BHs
after the discovery of GWs.

\citet{Hoyle_Lyttleton39}
considered the effect of the ISM accretion on the Sun's radiation 
for explaining changes in terrestrial climate.
\citet{Bondi_Hoyle44}
investigated the accretion in detail including the effect of perturbations.
\citet{Bondi52} included the pressure effects
to complete the Bondi-Hoyle formula.

\citet{Zel'dovich64} and \citet{Salpeter64}
suggested accretion onto a BH as an important source of radiation.
\citet{Shvartsman71}
treated both the fluid dynamics and radiative processes
by employing nonrelativistic approximations.
\citet{Michel72} considered the general-relativistic version of 
the Bondi-Hoyle accretion.
\citet{Shapiro73} made a general-relativistic treatment
of both the fluid mechanics and radiative processes.
\citet{Meszaros75} included the heating due to
dissipation of turbulence and magnetic reconnection,
and the radiation from cosmic rays captured with the accretion.

The formation of a disk around a massive BH is considered by
\citet{Lynden-Bell69}.
\citet{Pringle_Rees72} 
developed the accretion disk model.
\citet{Shakura_Sunyaev73} completed
the standard disk model.
\citet{Novikov_Thorne73}
made a general-relativistic version of the standard disk model.

\citet{Ipser_Price77,Ipser_Price82}
carried out detail calculations of radiation from Galactic massive BHs 
that are spherically accreting from the ISM.
\citet{Grindlay78} discussed X-ray limits on Galactic BHs 
that are spherically accreting from the ISM.
\citet{Carr79} placed a limit on BHs 
with the background radiation density.
\citet{McDowell85}
suggested that molecular clouds offer high density
for spherically accreting BHs to produce detectable fluxes.
\citet{Campana_Pardi93}
estimated the number of BHs in the molecular clouds.
\citet{Heckler_Kolb96} proposed a search strategy for 
isolated stellar mass BHs in the solar neighborhood using optical surveys.
\citet{Chisholm+03}
gave possible BH candidates using optical and X-ray surveys.

Modern estimates for radiation from isolated stellar mass BHs
began after
\citet{Fujita+98}
adopted the advection-dominated accretion flow (ADAF) model
and \citet{Armitage_Natarajan99}
considered a jet from an isolated accreting BH,
although they do not estimate the luminosity function, the total power,
and the acceleration energy.
They did not also consider the MAD state.

\citet{Agol_Kamionkowski02}
calculated the luminosity function of isolated accreting BHs,
although they assumed a constant efficiency of the disk emission
and do not consider a BH jet.

\citet{Barkov+12} also considered 
a jet from an isolated accreting BH,
although they do not consider the luminosity function
and hence their estimate on the total power is not correct.
They did not also consider the MAD state.

\citet{Fender+13}
calculated the X-ray luminosity distribution and 
suggested a discrepancy 
between the theoretical expectation and the hard X-ray surveys,
although their prescription for radiatively inefficient accretion
is very simple.

\citet{Nakamura+16} considered the Bondi-Hoyle accretion
for the optical counterparts of nearby BH mergers.


\bsp	
\label{lastpage}
\end{document}